\documentclass[orivec,10pt]{llncs}
\usepackage{smallsec}
\usepackage{amssymb}
\usepackage{amsmath}
\usepackage{paralist}
\usepackage{MnSymbol}
\usepackage{cite}
\usepackage{subfigure}
\usepackage{graphicx}
\usepackage{xspace}
\usepackage{wrapfig}
\usepackage{hyperref}

\begin{document}
\title{Synthesising Executable Gene Regulatory Networks from
Single-cell Gene Expression Data}

\author{
Jasmin Fisher\inst{1,2}
\and
Ali Sinan K\"oksal\inst{3}
\and
Nir Piterman\inst{4}
\and
Steven Woodhouse\inst{1}
}
\institute{
University of Cambridge, UK
\and
Microsoft Research Cambridge, UK
\and
University of California, Berkeley, USA
\and
University of Leicester, UK
}

\maketitle

\begin{abstract}
Recent experimental advances in biology allow researchers to obtain
gene expression profiles at single-cell resolution over hundreds, or even
thousands of cells at once. These single-cell measurements provide
snapshots 
of the states of the cells that make up a tissue, instead of the
population–-level averages provided by conventional high-throughput
experiments.
This new data therefore provides an exciting opportunity for
computational modelling. In this paper we introduce the idea of viewing
single-cell gene expression profiles as states of an asynchronous Boolean
network, and frame model inference as the problem of reconstructing a
Boolean network from its state space. 
We then give a scalable algorithm to solve this synthesis problem.
We apply our technique to both simulated and real data. 
We first apply our technique to data simulated from a well established
model of common myeloid progenitor differentiation. 
We show that our technique is able to recover the original Boolean network
rules.
We then apply our technique to a large dataset taken during embryonic
development containing thousands of cell measurements. 
Our technique synthesises matching Boolean networks, and analysis of
these models yields new predictions about blood development which our
experimental collaborators were able to verify.

\end{abstract}

\section{Introduction}
As biological data becomes more accurate and becomes available in
larger volumes, researchers are increasingly adopting concepts from
computer science to the modelling and analysis of living systems.
Formal methods have been successfully applied to gain insights into
biological processes and to direct the design of new experiments
\cite{Cook2011, Claessen2013, Koksal2013, Cook2014}.  New single-cell
resolution gene expression measurement technology provides an exciting
opportunity for modelling biological systems at the cellular
level. Single-cell gene expression profiles provide a snapshot of the
true states that cells can reach in the real experimental system, a
level of detail which has not been available before
\cite{Moignard2013, Pina2012}. A major challenge for researchers is to
move beyond established methods for the analysis of population data,
to new techniques that take advantage of single-cell resolution
data \cite{Moignard2014T}.

Uncovering and understanding the gene regulatory networks (GRNs) which
underlie the behaviour of stem and progenitor cells is a central issue
in molecular cell biology. These GRNs control the self-renewal and
differentiation capabilities of the stem cells that maintain adult
tissues, and become perturbed in diseases such as cancer. They also
specify the complex developmental processes that lead to the initial
formation of tissues in the embryo.  Understanding how to effectively
control GRNs can lead to important insights for the programmed
generation of clinically-relevant cell types important for
regenerative medicine, as well as into the design of molecular
therapies to target cancerous cells.

Biological systems can be modelled at different levels of abstraction.
At a molecular level, the biochemical events which occur inside a cell
can be captured by stochastic processes, given by chemical master
equations \cite{Wilkinson2012}. These chemical events are
fundamentally stochastic, driven by random fluctuations of molecules
present at low concentrations and by Brownian motion.  Asynchronous
Boolean networks abstract away details of transcription, translation
and molecular binding reactions and represent the status of each
modelled substance as either active (on) or inactive (off), while
using non-determinism to capture different options that arise from
stochastic behaviour~\cite{Krumsiek2011,Garg2008, Zheng2013}.
In the cell, gene activity is controlled by combinatorial logic in
which proteins called transcription factors cooperate to physically
bind to a regulatory DNA region of a gene and trigger (or inhibit) its
transcription. Target genes may in turn code for transcription
factors, forming a complex GRN. Asynchronous Boolean networks are
particularly well suited to modelling GRNs because the combinatorial
logic regulating gene activity can be expressed as a Boolean
function. For example, gene X may be activated by either the presence
of gene A or by the presence of both genes B and C. The presence of a
repressor D may prevent X from becoming triggered by the presence of these
activating genes.  When modelling the differentiation of a cell using an
asynchronous Boolean network, dynamics proceed by a series of single--gene
changes.  Mature, differentiated cell types correspond to stable attractor
states of the model.

Predictions about the modes of interaction between genes resulting
from computational analysis can be tested experimentally through a
range of assays. For example, if analysis of a model predicts that
gene X is activated by gene A, a ChIP (Chromatin ImmunoPrecipitation)
assay can be used to assess whether the protein coded for by A binds
to a regulatory region of X. Then, perturbations 
which prevent the binding of A to this region can be introduced, and
the effect that this has on the expression of X can be examined.

State--space analyses of hand--built asynchronous Boolean network models based on
literature--derived gene regulatory interactions have been successfully
applied to model cell fate decisions, and to reproduce known experimental
results (e.g.,~\cite{Krumsiek2011, Bonzanni2013, Kazemzadeh2012}).
Here we address the problem of automatically constructing such models directly from data.
If we think of single-cell gene expression profiles as the state space of an asynchronous
Boolean network, can we identify the underlying gene regulatory logic that could have
generated this data?

We encode the matching of an asynchronous Boolean network to a
state space as a synthesis problem and use constraint (satisfiability)
solving techniques for answering the synthesis problem.
The synthesised network has to match the data in two aspects.
First, the resulting network should try to minimise transitions to
expression points that are not part of the sampled data.
Second, the resulting network should allow for a progression through
the state space in a way that matches the flow of time through the
different experiments that produced the data.
A direct encoding of this problem into a satisfiability problem does
not scale well.
We suggest a modular search that handles parts of the state space and
the network and does not need to reason about the entire network at
once.
We consider two test cases.
First, we try to reconstruct an existing asynchronous Boolean network
from its state space.
We are able to reconstruct Boolean rules from the original network.
Second, we apply our technique to experimental data derived from blood
cell development.
The network that is produced by our technique matches known
dependencies and suggests interesting novel predictions.
Some of these predictions were validated by our collaborators.

This paper describes the algorithm that we used to obtain the
results in a recently published biological paper on a single-cell
resolution study of embryonic blood development~\cite{Moignard2015}.
The biological paper
includes full details of the experiment
that generated the data, and the biological validation of
our resulting synthesised model. Here, we cover the algorithmic
aspects of our 
method.

\section{Biological Motivation}
\label{sec:motivation}

\begin{figure}[bt]
  \begin{center}
    \includegraphics[width=0.4\textwidth]{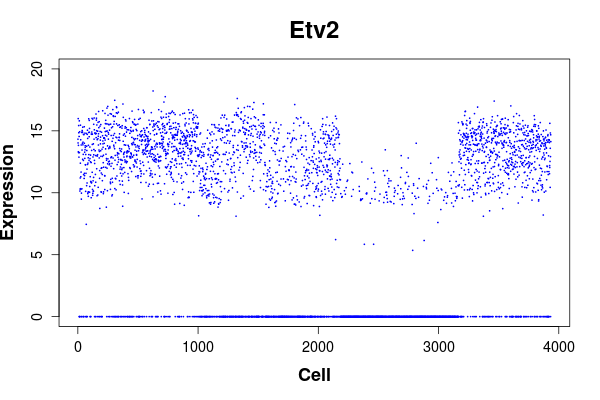}
    \includegraphics[width=0.4\textwidth]{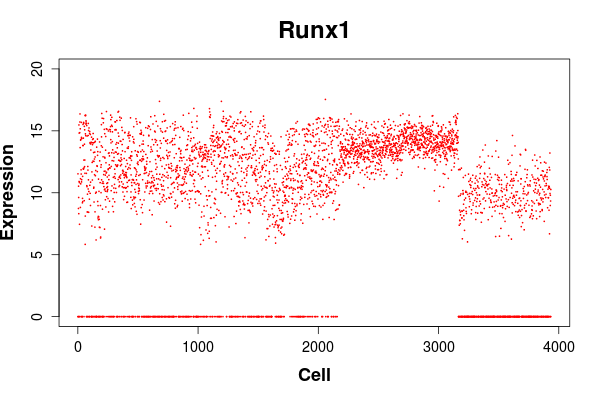}
  \end{center}
  \vspace*{-6mm}
\caption{\label{fig:single cell}Single--cell gene expression measurements for two genes, in 3934 cells.}
  \vspace*{-5mm}
\end{figure}

\begin{figure}[bt]
  \begin{center}
   \includegraphics[width=0.5\textwidth]{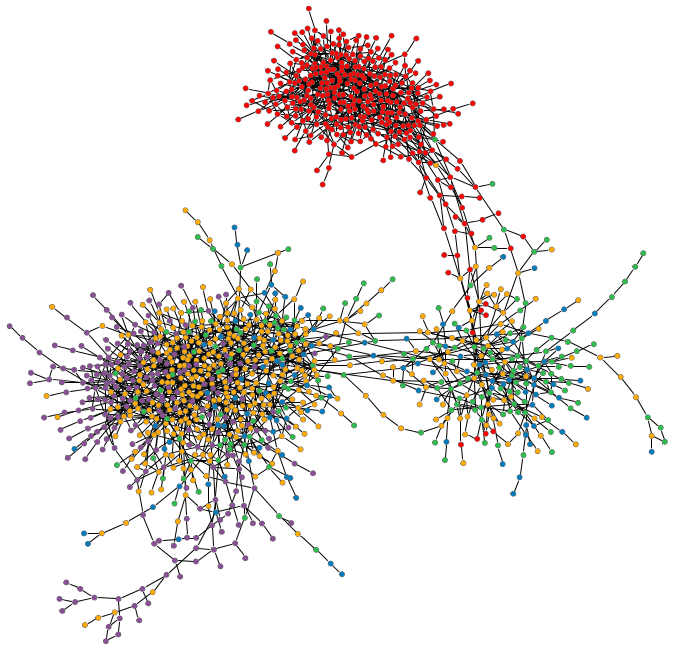}
   \end{center}
   \vspace*{-6mm}
\caption{\label{fig:chicken}State graph. Node colours correspond to the time point at which a state was measured. States from the earliest of the time points are coloured blue, and states from the last time
  point are coloured red.}
   \vspace*{-7mm}
\end{figure}

Single-cell gene expression experiments produce gene expression
profiles for individually measured cells. Each of these gene expression
profiles is a vector where each element gives the level of expression of
one gene in that cell. Figure~\ref{fig:single cell} plots
the level of the genes \textit{Etv2} and \textit{Runx1} over 3934 cells.

Our experimental collaborators performed such gene expression profiling on
five batches of cells taken from four sequential developmental time
points of a mouse embryo.  For each time point, the experiment aimed to capture
every cell with the potential to develop into a blood cell, providing a comprehensive
single--cell resolution picture of the developmental timecourse of blood development.
This resulted in a data set of 3934 cell measurements.
Full details of this experiment and our analysis can be found in
\cite{Moignard2015}.
This data set is the first of its kind, attempting to capture an entire tissue's
worth of progenitor cells across a developmental time course. This level of coverage
of the potential cell state space is required for our approach to accurately recover
gene regulatory networks, and requires the measurement of thousands of cell profiles.
Later we will introduce a synthetic data set of a few hundred cell states in order to
illustrate how our approach works, but we would like to stress that to be usable on real
experimental data our algorithm needs to be able to scale thousands of cell states.

For each of 3934 cells, the level of expression of 33 transcription
factor genes was measured. Expression levels are non-negative real numbers, where the value $0$
indicates that the given gene is unexpressed in the cell (see Figure~\ref{fig:single cell}).

The key idea introduced in this paper is to view this gene-expression
data as a sample from the state-space of an asynchronous Boolean
network.  In the past, manually curated Boolean networks have been
successfully used to recapitulate experimental
results~\cite{Bonzanni2013,Krumsiek2011, Kazemzadeh2012}.  Such
Boolean networks were hand--constructed from biological knowledge that has
accumulated in the literature over many years.  Here, we aim to
produce such Boolean networks automatically, directly from gene expression data, by employing synthesis
techniques.  We aim to produce a Boolean network that can explain the
data and can be used to inform biological experiments for uncovering
the nature of gene regulatory networks in real biological systems.

In order to convert the data into a format that can be viewed as a Boolean
network state space, we first discretise expression values to binary,
assigning the value 1 to all non-zero gene expression measurements.
A value of zero corresponds to the discovery threshold of the equipment used to
produce the data.
Discretising the 3934 expression profiles in this way yields 3070 unique
binary states, where every state is a vector of 33 Boolean values corresponding to
the activation/inactivation level of each of 33 genes in a given cell.
In an asynchronous Boolean network, transitions correspond to the
change of value of a single variable.
Hence, we next look for pairs of states that differ by only one gene (that is, 
the Hamming distance between the two vectors is 1).
An analysis of the strongly-connected components of this graph shows
that one strongly connected component contains 44\% of the states.
We note that in a random sample of 3934 elements from a space of
$2^{33}$, the chance of seeing repeats or neighbours with
Hamming--distance 1 is negligible.

A plot of the graph of the largest strongly connected component is
given in Figure~\ref{fig:chicken}.  We add an edge for every
Hamming--distance 1 pair and cluster together highly connected nodes.
The colours of nodes correspond to the developmental time the
measurements was taken. Note that there is
a clear separation between the earliest developmental time point
and the latest one.  This representation 
already suggests a clear change of states over the development of the
embryo, with separate clusters identifiable and obvious fate
transitions between clusters.

We wish to find an asynchronous Boolean network that matches this
graph.
For that we impose several restrictions on the Boolean network.
Connections between states correspond to a change in the value
of one gene, however, we do not know the direction of the change.
Thus, we search simultaneously for directions and update functions of
the different genes that satisfy the following two conditions:
states from the earliest developmental time point
should be able to evolve, through a series of single--gene transitions, to the states from 
the latest developmental time point.
Secondly, the 
update functions must minimise the number of transitions that
lead to additional, unobserved states, that were not measured in the experiment.

\section{Example: Reconstructing an ABN from its State Space}
\label{sec:example}

\begin{wrapfigure}[11]{r}[-20pt]{0.4\textwidth}
  \vspace*{-8mm}
  {\scriptsize
    \begin{tabular}{ | l | l |}
    \hline
    \textbf{Gene} & \textbf{Update function} \\ \hline
    Gata2 & $\textit{Gata2} \land \lnot (\textit{Pu.1} \lor (\textit{Gata1} \land \textit{Fog1}))$ \\ \hline
    Gata1 & $(\textit{Gata1} \lor \textit{Gata2} \lor \textit{Fli1}) \land \lnot \textit{Pu.1}$ \\ \hline
    Fog1 & $\textit{Gata1}$ \\ \hline
    EKLF & $\textit{Gata1} \land \lnot \textit{Fli1}$  \\ \hline
    Fli1 & $\textit{Gata1} \land \lnot \textit{EKLF}$  \\ \hline
    Scl & $\textit{Gata1} \land \lnot \textit{Pu.1}$ \\ \hline
    Cebpa & $\textit{Cebpa} \land \lnot (\textit{Scl} \lor
    (\textit{Fog1} \land \textit{Gata1}))$
\\ \hline
    Pu.1 & $(\textit{Cebpa} \lor \textit{Pu.1}) \land \lnot (\textit{Gata1} \lor \textit{Gata2})$ \\ \hline
    cJun & $\textit{Pu.1} \land \lnot \textit{Gfi1}$ \\ \hline
    EgrNab & $(\textit{Pu.1} \land \textit{cJun}) \land \lnot \textit{Gfi1}$ \\ \hline
    Gfi1 & $\textit{Cebpa} \land \lnot \textit{EgrNab}$ \\ \hline
    \end{tabular}
  }
  \vspace*{-4mm}
  \caption{\label{fig:bn}Boolean update functions for a manually curated
    network.}
  \vspace*{-5mm}
\end{wrapfigure}

We first illustrate our synthesis method using an example. We take an
existing Boolean network, construct its associated state space, and
then use this state space as input to our synthesis method in order to
try to reconstruct the Boolean network that we started with.

Krumsiek {\em et.~al.}~ introduce a Boolean network model of the core
regulatory network active in common myeloid progenitor cells
\cite{Krumsiek2011}.  Their network is based upon a comprehensive
literature survey.  It includes a set of 11 Boolean variables
(corresponding to genes) and a Boolean update function for each
variable (Figure~\ref{fig:bn}).\footnote{The function of \textit{Cebpa} is
  modified from that in \cite{Krumsiek2011} to match the format we
  assume.}  The model is given a well-defined initial starting state,
representing the expression profile of the common myeloid progenitor,
and computational analysis reveals an acyclic, hierarchical state
space of 214 states with four stable state attractors (Figure~\ref{fig:cmp
  graph}).

These stable
attractors are in agreement with experimental expression profiles
of megakaryocytes, erythrocytes, granulocytes and monocytes; four of
the mature myeloid cell types that develop from common myeloid
progenitors.

\begin{figure}[bt]
  \begin{minipage}[b]{0.55\linewidth}
    \centering
    \includegraphics[width=\textwidth]{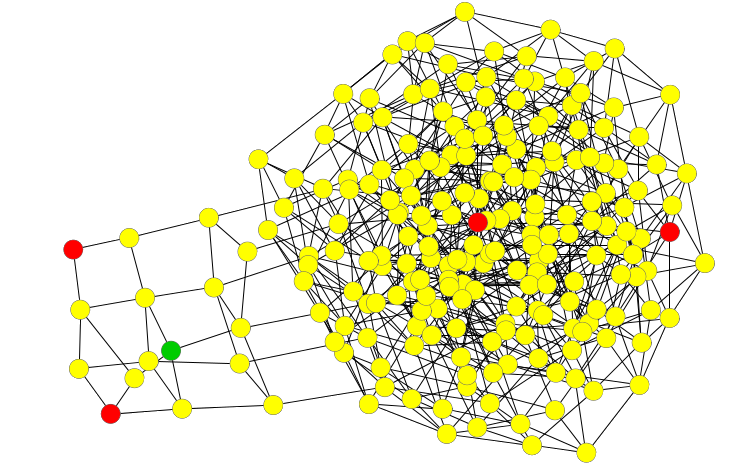}
    \caption{\label{fig:cmp graph}Boolean network state space. Initial
      state is coloured green, stable states red.}
    \vspace*{-6mm}
  \end{minipage}
  \hspace{0.01\textwidth}
  \begin{minipage}[b]{0.45\linewidth}
    \centering
    \includegraphics[width=\textwidth]{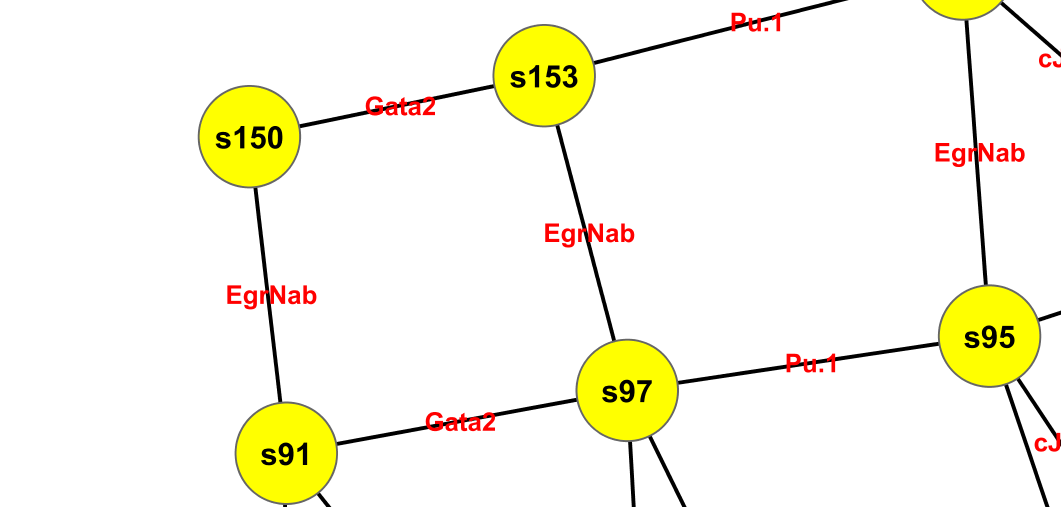}
    \caption{\label{fig:cmp zoom}Close--up of Boolean network state
      space.}
    \vspace*{-6mm}
  \end{minipage}
\end{figure}

We treat the state space of this Boolean network as we would treat
experimental data, forgetting all directionality information, and
connecting all states which differ in the expression of only one gene
by an undirected edge (Figures~\ref{fig:cmp graph} and~\ref{fig:cmp
  zoom}, where each edge is labelled with the single gene that changes
in value between the states it connects).
We would now like to reconstruct the Boolean network given in
Figure~\ref{fig:bn} from this undirected state space. 

For each gene, we would like to assign a direction to each of its 
labelled edges (or decide that it does not exist), in a way that is
compatible with a Boolean update function. 
For example, in Figure~\ref{fig:cmp zoom}, we may orient the
\textit{Pu.1}-labelled edge between states 97 and 95 in the
direction $s_{97} \to s_{95}$, in the direction
$s_{95} \to s_{97}$, or decide that this is not a possible update.
We also allow the edge to be directed in both directions.
If $s_{97} \to s_{95}$, we want a Boolean update function
$u_\textit{Pu.1}$ that takes state $s_{97}$ to state $s_{95}$.
Since there is no \emph{Pu.1}--labelled edge leaving state $s_{150}$,
we can also add the constraint that $u_\textit{Pu.1}$ takes $s_{150}$
to $s_{150}$.

We also add reachability constraints that restrict which edges are
included and their orientation.
Since the state space was constructed starting from a well-defined initial state,
we would like to enforce the constraint that each non-initial state ought to be
reachable by some directed path from the initial state. Since cell development
proceeds hierarchically and unidirectionally, we favour short paths over long paths.
This eliminates routes that seem biologically implausible,
for example routes that cross a fate transition and then return to
where they began. It also reduces the space of paths we have to search
through.
By increasing the lengths of allowed paths, we can increase the number
of considered solutions.

\begin{figure}[bt]
  \begin{center}
  {\scriptsize
    \begin{tabular}{ | l | l | l |}
    \hline
    \textbf{Gene} & \textbf{Synthesised update functions} & \textbf{Comments} \\ \hline
    Gata2 & $\textit{Gata2} \land \lnot (\textit{Fog1} \lor \textit{Pu.1})$ &  \\
          & $\textit{Gata2} \land \lnot (\textit{Fog1} \lor (\textit{Pu.1} \land \textit{Cebpa}))$ &  \\
          & $\textit{Gata2} \land \lnot (\textit{Fog1} \lor (\textit{Pu.1} \land \textit{Gata2}))$ &  \\
          & $\textit{Gata2} \land \lnot (\textit{Gata2} \land (\textit{Pu.1} \lor \textit{Fog1})$ &  \\
          & \textbf{$\textit{Gata2} \land \lnot (\textit{Pu.1} \lor (\textit{Gata1} \land \textit{Fog1}))$} &  \\
          & $\textit{Gata2} \land \lnot (\textit{Pu.1} \lor (\textit{Gata2} \land \textit{Fog1}))$ &  \\ \hline
    Gata1 & $(\textit{Gata1} \lor \textit{Cebpa}) \land \lnot \textit{Pu.1}$ &  \\
          & $(\textit{Gata2} \lor \textit{Fog1}) \land \lnot \textit{Pu.1}$ &  \\
          & $(\textit{Gata1} \lor \textit{Gata2}) \land \lnot \textit{Pu.1}$ &  \\
          & \textbf{$(\textit{Gata1} \lor \textit{Gata2} \lor \textit{Fli1}) \land \lnot \textit{Pu.1}$} &  \\
          & Other functions of the form $(X \lor Y \lor Z) \land \lnot \textit{Pu.1}$ &  \\ \hline
    Fog1 & \textbf{$\textit{Gata1}$} & Unique \\ \hline
    EKLF & \textbf{$\textit{Gata1} \land \lnot \textit{Fli1}$} & Unique \\ \hline
    Fli1 & \textbf{$\textit{Gata1} \land \lnot \textit{EKLF}$} & Unique \\ \hline
    Scl & $\textit{Gata1}$ &  \\
        & \textbf{$\textit{Gata1} \land \lnot \textit{Pu.1}$} &  \\ \hline
    Cebpa & $\textit{Cebpa} \land \lnot (\textit{Fog1} \lor \textit{Scl})$ &  \\
      & $\textit{Cebpa} \land \lnot (\textit{Cebpa} \land (\textit{Scl} \lor \textit{Fog1}))$ &  \\
      & $\textit{Cebpa} \land \lnot (\textit{Fog1} \land (\textit{Scl} \lor \textit{Cebpa}))$ &  \\
      & $\textit{Cebpa} \land \lnot (\textit{Fog1} \lor (\textit{Scl} \land \textit{Gata1}))$ &  \\
      & $\textit{Cebpa} \land \lnot (\textit{Fog1} \lor (\textit{Scl} \land \textit{Gata2}))$ &  \\
      & $\textit{Cebpa} \land \lnot (\textit{Gata1} \land (\textit{Fog1} \lor \textit{Scl})$ & \\
      & $\textit{Cebpa} \land \lnot (\textit{Scl} \lor (\textit{Fog1} \land \textit{Cebpa})$ &  \\
      & \textbf{$\textit{Cebpa} \land \lnot (\textit{Scl} \lor (\textit{Fog1} \land \textit{Gata1})$} &  \\ \hline
    Pu.1 & $\textit{Pu.1} \land \lnot \textit{Gata2}$ &  \\
      & $(\textit{Pu.1} \land \textit{Cebpa}) \land \lnot \textit{Gata2}$ &  \\
      & $\textit{Pu.1} \land \lnot (\textit{Gata1} \lor \textit{Gata2})$ &  \\ 
      & Other functions of the form $\textit{Pu.1} \land \lnot (\textit{Gata2} \lor X)$ &  \\
      & $\textit{Pu.1} \land \lnot (\textit{Gata2} \land \textit{Cepba})$ &  \\
      & $\textit{Pu.1} \land \lnot (\textit{Gata2} \land \textit{Pu.1})$ &  \\
      & $\textit{Cebpa} \land \lnot (\textit{Gata1} \lor \textit{Gata2})$ &  \\
      & $\textit{Cebpa} \land \lnot (\textit{Gata2} \lor \textit{Fog1})$ &  \\
      & \textbf{$(\textit{Cebpa} \lor \textit{Pu.1}) \land \lnot (\textit{Gata1} \lor \textit{Gata2})$} &  \\
      & $(\textit{Cebpa} \land \textit{Pu.1}) \land \lnot (\textit{Gata1} \lor \textit{Gata2})$ &  \\
      & Other functions of the form $(\textit{Cebpa} \lor X) \land \lnot (\textit{Gata2} \lor Y)$ &  \\
      & Other functions of the form $(\textit{Pu.1} \lor X) \land \lnot (\textit{Gata2} \lor Y)$ &  \\
      & Other functions of the form $(\textit{Cebpa} \land \textit{Pu.1}) \land \lnot (\textit{Gata2} \lor X)$ &   \\ \hline
    cJun & \textbf{$\textit{Pu.1} \land \lnot \textit{Gfi1}$} & Unique \\ \hline
    EgrNab & $(\textit{cJun} \lor \textit{Gata1}) \land \lnot \textit{Gfi1}$ & Incorrect with shortest paths \\ \hline
    Gfi1 & \textbf{$\textit{Cebpa} \land \lnot \textit{EgrNab}$} & Unique \\ \hline
    \end{tabular}
    }
  \end{center}
  \vspace*{-5mm}
  \caption{\label{fig:results}Synthesised update functions.}
  \vspace*{-7mm}
\end{figure}

The results of applying our technique
are shown in Figure ~\ref{fig:results}.
The method reconstructs the Boolean update functions for
all but one gene (\textit{EgrNab}), in some cases 
uniquely identifying the original function.
We note that when multiple solutions are found for an update function,
these solutions, while not exact, all provide useful regulatory
information that could be verified experimentally. For
example, both solutions for \textit{Scl} successfully predict
\textit{Scl}'s activation by \textit{Gata1}, although one of the two
solutions omits its repression by \textit{Pu.1}.

\newpage

\section{Background to Asynchronous Boolean Networks}
An \textit{asynchronous Boolean network} (ABN) is $B(V, U)$, where $V
= \{v_1, v_2, \dots, v_n\}$ is a set of \textit{variables}, and $U=
\{u_1, u_2,\dots, u_n\}$ is a set of Boolean \textit{update
  functions}.  For every $u_i \in U$ we have $u_i : \{0,1\}^n \to
\{0,1\}$ associated with variable $v_i$.  A \emph{state} of the system
is a map $s : V \to \{0,1\}$. We say that an update function $u_i$ is
\textit{enabled} at state $s$ if $u_i(s) \neq s(v_i)$, i.e. applying
the update function $u_i$ to state $s$ changes the value of variable
$v_i$.

State $s'=(d'_1,d'_2,\dots, d'_n)$ is a \textit{successor} of state $s
= (d_1, d_2, \dots, d_i, \dots, d_n)$ if for some $i$ we have $u_i$ is
enabled, $d'_i=u_i(s)$, and for all $j\neq i$ we have $d'_j=d_j$.
That is, we get to the next state $s'$, by non-deterministically
selecting an enabled update function $u_i$ and updating the value of
the associated variable: $s' = (d_1, d_2, \dots, u_i(d_i), \dots,
d_n)$.  If no update function is enabled, the system remains in its
current, stable, state, where it will remain: $s' = s$.

An ABN induces a labelled transition system $T=(N,R)$, where $N$ is
the set of $2^n$ states of the ABN, and $R\subseteq N\times V \times
N$ is the successor relation.
Each transition $(s_1,v_i,s_2)$ is labelled with the variable
$v_i$ such that $s_1(v_i)\neq s_2(v_i)$.

The \textit{undirected state space} of an ABN is an undirected graph
$S=(N,E)$, where each vertex $n \in N$ is uniquely labelled with a
state $s$ of the Boolean network, and there is an edge $\{s_1,s_2\}\in
E$ iff $s_1$ and $s_2$ differ in the value of exactly one variable,
$v$. The edge $\{s_1,s_2\}$ is labelled with $v$.
In general, an undirected state space does not have to include all
$2^n$ states induced by a Boolean network.

An ABN $B(V,U)$ \emph{induces} a \textit{directed state space} on an
undirected state space $S=(N,E)$.
Consider the transition system $T=(2^V,R)$ of $B(U,V)$.
Then, the induced directed state space is $S'=(N,A)$,
where $(s_1,s_2)\in A$ implies that there is a variable $v_i$
such that $(s_1,v_i,s_2)\in R$. 
We say that $(s_1,s_2)$ is \textit{compatible} with $u_{i}$, if
$s_2(v_i)=u_i(s_1)$, and for every $j\neq i$ we have $s_2(v_j)=s_1(v_j)$.

\section{Formal Definition of the Problem}
\label{sec:formal def}
Our synthesis problem can be stated as follows: we are given an
undirected state space $S$ over a given set of variables $V$.  We
would like to extract a set of Boolean update functions that induce a
directed state space from $S$ such that each of the states in $S$ are
reachable from a given set of initial states. We also want to ensure
that no additional, undesired states not in $S$ are reachable, by
ruling out transitions which `exit' the state space.

More formally, we are given a set of variables
$V=\{v_1,v_2,\dots,v_n\}$, an undirected state space $S = (N, E)$
over $V$, and a set $I\subseteq N$ of
\textit{initial} vertices.

We would like to find an update function $u_i : \{0,1\}^n \to \{0,1\}$
for each variable $v_i \in V$, such that the following conditions
hold.
Let $U=\{u_i ~|~ v_i\in V\}$ be the set of update functions.

\begin{compactenum}
\item 
  Every non-initial vertex $s\in N - I$ is reachable from some initial
  vertex $s_i\in I$ by a directed path in the directed state space
  induced by $B(V,U)$ on $S$.
\item 
  For every variable $v_i \in V$, let $N_i$ be the set of states
  without an outgoing $v_i$-labelled arc.  For every $i$ we require
  that for each $s\in N_i$, $u_i(s)=s(v_i)$.
\end{compactenum}

\subsection{Generalising the Definition to Partial Data}
\label{sec:formal def partial}
Since we intend to apply our method in an experimental setting, where
we only have an incomplete sample from the possible states of the
system, we relax this definition to extend it to partial data. Instead
of requiring that \textit{every} state is reachable from those initial
states that we have measured, we only require that a set of
\textit{final} states are reachable.  Instead of requiring that every
undesired transition is ruled out, we seek to maximise the number of
such transitions which are eliminated.
This is formally stated next.

As before, we are given a set of variables $V=\{v_1,v_2,\dots,v_n\}$, an undirected state space $S = (N, E)$
over $V$, and a designated set $I\subseteq N$ of \textit{initial}
vertices. In addition, we are given a designated set $F\subseteq N$ of \textit{final}
vertices, along with a \textit{threshold} $t_i$ for each variable $v_i \in V$.
The threshold $t_i$ specifies how many undesired transitions must be ruled out.

We would like to find an update function $u_i : \{0,1\}^n \to \{0,1\}$
for each variable $v_i \in V$, such that the following conditions
hold.
Let $U=\{u_i ~|~ v_i\in V\}$ be the set of update functions.

\begin{compactenum}
\item 
  Every final vertex $s_f\in F$ is reachable from some initial vertex
  $s_i\in I$ by a directed path in the directed state space induced by
  $B(V,U)$ on $S$.
\item
  For every variable $v_i \in V$, let $N_i$ be the set of states
  without an outgoing $v_i$-labelled arc.  For every $i$ the number of
  states $s\in N_i$ such that $u_i(s)=s(v_i)$ is greater or equal to
  $t_i$.
\end{compactenum}

In the remainder of the text, we refer to condition 1 as the
\textit{reachability condition} and condition 2 as the
\textit{threshold condition}.

We restrict the search to update functions of the form
$
f_1 \wedge \neg f_2
$,
where $f_i$ is a monotone Boolean formula.
The inputs to $f_1$ are the activating inputs to the gene and the
inputs to $f_2$ are the the repressing inputs. This restriction was chosen after
discussion with biologist colleagues and consultation of the literature (e.g.,~\cite{Krumsiek2011, Bonzanni2013}).

\section{A Direct Encoding}
\label{sec:direct}
We start with a direct encoding of the search for a matching Boolean
network.
The search is parameterised by the shape of update functions (how many
activators and how many repressors each variable has), 
the length of paths from initial states to final states, and the
thresholds for each variable.
By increasing the first two parameters and decreasing the last we can explore
all possible Boolean networks.

\subsection{Possible Update Functions}
In order to represent the Boolean update function for gene $v_i$, $u_i
= f_1 \wedge \neg f_2$, we use a bitvector encoding.
We represent the
Boolean formula $f_j$ by a set of bitvectors, $\{a_1, a_2, \dots
a_n\}$, $a_j \in V \cup \{\lor, \land\}$, where each bitvector $a_i$
represents a variable or a Boolean operator, and solutions take the
form of a binary tree.
For example, the formula $v_1 \land (v_2 \lor
v_3)$ is represented by the solution $a_1 = \land, a_2 = \lor, a_3 = v_1, a_4 =
v_2, a_5 = v_3$. We restrict the syntactic form
of possible update functions so that each variable appears only once,
and each possible function has one canonical representation. For
example, the function $(v_1 \land (v_2 \lor v_3))$ is included in our
search space while $(v_1 \land v_2) \lor (v_1 \land v_3)$ is not.  We
search for functions up to a maximum number of activators, $A_i$, and a
maximum number of repressors, $R_i$.

To encode the application of function $u_i$ to a state $s$, $u_i(s)$,
we add implications which unwrap the bitvector encoding of $u_i$ to
the constituent variables and logical operators; substituting values,
$s(v_j)$, for variables, $v_j$, and directly mapping operations to logical
constraints in the Boolean satisfiability formula. For example, the
application of the function $(v_1 \lor v_2) \land \lnot v_3$ to the state
$s_1$ is mapped to $(s_1(v_1) \lor s_1(v_2)) \land \lnot s_1(v_3)$.

\subsection{Ensuring Reachability}
To enforce the global reachability condition we consider all of the underlying
directed edges in the undirected state space $S = (N, E)$, and their associated
single--gene transitions.

Recall that we require every final vertex to be reachable from some
initial vertex by a directed path in the directed state space induced
on $S$ by the Boolean network. That is, we require that every final
vertex is reachable by a directed path, and that every $v_j$-labelled edge
along this path is compatible with its associated update function, $u_j$.

To enforce this we add constraints that track the compatibility of edges with
update functions and define reachability recursively. We consider reachability by
paths up to a maximum length: recall that we consider shorter paths to be more
biologically likely. By iteratively increasing the length
of the paths considered, we can obtain all satisfying models.

We introduce a pair of Boolean variables $e_{ij}, e_{ji}$ for each
$v_i$-labelled undirected edge $\{s_i,s_j\} \in E$, which track the
value of the application of $u_i$ to $s_i$ and to $s_j$ (and the
compatibility of the underlying directed edges $(s_i, s_j)$
and $(s_j, s_i)$ with $u_i$).
$e_{ij}$ is
true iff $u_i(s_i)=s_j(v)$.

We introduce an integer given by a bitvector encoding, $r_n$, for each
node $n \in N$.  Bitvector $r_n$ encodes the fact that node $n$ is
reachable from an initial node in $r_n$ steps, up to some maximum
encodable value $2^{|r_n|}-1$.
Bitvector $r_n$ is given a value of -1 to
indicate that $n$ is not reachable in this maximum number of steps.

Reachability is then defined inductively:

\begin{compactenum}
\item 
  Initial nodes are reachable in zero steps: for every $i \in I$, $r_i = 0$.
\item
  A non--initial node $s_i$ is reachable in $M$ steps if there is a compatible incoming
  edge $(s_j, s_i)$ from another node $s_j$, and $s_j$ is itself reachable in fewer than $M$ steps.
  That is, for every $n = s_j \in N - I$ and $m = s_i \in N$ such that
  $\{s_i,s_j\} \in E$ we have $e_{ij} \rightarrow r_m < r_n$.
  We also have that non--initial nodes cannot be reached in zero steps:
  For every $n \in N - I$, $r_n = -1 \lor r_n > 0$.
\end{compactenum}

Finally, we add a constraint that every final node $n \in F$ is
reachable from some initial node: $r_n \neq - 1$.

\subsection{Enforcing the Threshold Condition}
We enforce the threshold condition for
each update function as follows.

Consider an update function $u_i:V\rightarrow \{0,1\}$.  We say that a
node $s\in N_i$ is {\em negatively matched} by $u_i$ if $u_i(s)=s(v_i)$.
That is, by using $u_i$ as the update function of variable $v_i$,
$u_i$ does not change the value of $v_i$ from node $s$.  We are
searching for an update function such that a maximum number of nodes
from $N_i$ are negatively matched.

We add a variable, $m_{is}$ for each node $s \in N_i$ to record whether
$u_i$ negatively matches $s$. We then add a constraint demanding that
the number of negatively matched nodes is greater than or equal to the
threshold:
%
$
\sum_{s \in N_i}m_{is} \geq t_i
$.

We search for satisfying assignments to the constraint variables
encoding the representation of the Boolean update functions $u_i$ for
all $v_i$ in $V$. The resulting synthesised Boolean network is the
combination of these update functions.

Unfortunately, in practice the direct encoding of the search does not scale to
handle our experimental data.  In the next section we suggest a compositional
way to solve the problem. 

\section{A Compositional Algorithm}
We now introduce our compositional algorithm, which scales better than
the direct encoding given above.  The problem of synthesising a
Boolean network from the data is partitioned to three
stages. Crucially, we avoid searching for a complete Boolean network
and consider parts of the network that can be constructed
independently.

\subsection{Pruning the Set of Possible Edges}
We start by building a directed graph from the given undirected state space
$S = (N, E)$, by considering which of the underlying directed edges in $E$
are compatible with some Boolean update function, and pruning those that
are not. We consider each underlying directed edge $(s_1, s_2)$ and $(s_2, s_1)$
of each of the $v_i$-labelled undirected edges $\{s_1, s_2\}$ in $E$ independently.

We pose a decision problem for each directed edge $(s_1, s_2)$:
whether there exists some Boolean update function $u_i$ satisfying the
threshold condition (condition 2, \ref{sec:formal def partial}) such that $u_i(s_1) =
s_2(v_i)$.  This is encoded as a Boolean satisfiability problem, adding
constraints to represent the encoding of the update function, the
threshold condition, and the evaluation of the function at the
specific edge under consideration.  We say that a satisfying function,
$u_i$, is \textit{compatible} with $(s_1, s_2)$.
Once a compatible function has been found, it can quickly be evaluated
outside the solver at other edges to try reduce the number of SAT queries we have to make.

After making a query for each edge, we are left with a directed
graph, which is the existential projection of all compatible update
functions for each of the variables $v \in V$.  We have eliminated
edges which have no compatible update function, and cannot participate
in the reachability condition. On the example data set from
Section~\ref{sec:example}, this step removes 18\% of the possible edges.

\subsection{Ensuring Reachability}
We now come to the only part of the algorithm that considers the edges
of all variables together, in order to enforce the global reachability
condition (condition 1, \ref{sec:formal def partial}).
This phase does not require the solving of a Boolean
satisfiability problem, and as a result is very efficient.

We construct, for each pair of initial nodes $i \in I$ and final nodes
$f \in F$, the shortest path $p_{if}$ from $i$ to $f$ in the directed
graph that was built in the previous phase of the algorithm.  These
paths can be computed via a breadth--first search.

Due to the edge pruning of the previous phase of the algorithm, if
there is no path to a final node $f$, this implies that there are no
satisfying models (at the given threshold and function size
parameters).  Otherwise, our reachability condition will be enforced
by fixing a set of directed edges $P_i$ for each variable $v_i \in V$
corresponding to these shortest paths. We will then require that the
update function we search for, $u_i$, is compatible with each of the
edges in $P_i$.

We choose, for each final node $f$, one path $p_f$ = $p_{if}$ from one
of the initial nodes $i$. By fixing this path, we ensure that $f$ is
reachable from an initial node.  We define $p_f|_i$ as the set of
$v_i$-labelled edges in the path $p_f$.  We define $P_i$, the
$v_i$-labelled edges which must be fixed to ensure reachability via
the chosen paths, as the the set of $v_i$-labelled edges in
$p_f$ for each final node $f$:

\begin{equation}
\label{eq:form of update functions third}
P_i = \bigcup_{f \in F}\{(s_1, s_2) ~|~ (s_1, s_2) \in p_f|_i\}
\end{equation}

By considering only the edges in $P_i$, we can search for an update
function for $v_i$ independently of all other variables, while
ensuring the global reachability condition holds.

\subsection{Final Update Functions}
\label{subsec:final update}
We can now search for the update function of variable $v_i$, $u_i$,
independently of all other variables.  We fix the $v_i$-labelled edges
computed in the previous phase and encode the search for $u_i$ as a
Boolean satisfiability problem.

As before we add constraints to encode the representation of $u_i$,
and to enforce the threshold condition.  We fix each of the
$v_i$-labelled edges $(s_1, s_2) \in P_i$ to establish reachability,
by adding a conjunction requiring that $u_i$ is compatible with each
of them: $u_i(s_1) = s_2(v_i)$.

We search for satisfying assignments of the constraint variables
encoding $u_i$, using an {\sc allsat} procedure to extract all
possible update functions for variable $v_i$.  This gives rise to a
set of update functions per variable and a set of Boolean
networks from the product of the set of update functions per
variable.

We note that this final phase of the algorithm can fail to find
update functions for a variable $v_i$, because there are no possible
update functions compatible with all of the path edges $P_i$ that were
computed in the previous phase.  That is, while each edge in $P_i$ is
individually compatible with some update function, there may be no update
function that is compatible with every edge in $P_i$.  In order to
cope with this limitation, we can extract the minimal unsatisifiable
core of the Boolean formula, and search for replacement paths that
exclude incompatible combinations of edges.  This step can be iterated
until satisfying solutions are found for all variables, or until no
path can be found, implying that there are no valid models.

By extending our search from the shortest paths between initial and
final node pairs in the directed graph to the $k$-shortest paths
between pairs and incremementally increasing $k$ \cite{Yen1971}, we
can increase the 
number of possible update functions that we consider. 
In the limit, we will obtain all satisfying models.

\begin{figure}[bt]
\begin{center}
\begin{tabular}{|l|c|c|c|c|}
  \hline
  \textbf{Data set} & \textbf{Genes} & \textbf{States} & \textbf{Direct (seconds)} & \textbf{Compositional (seconds)} \\
  \hline
  CMP (synthetic) & 11 & 214 & 25 & 77 \\
  Blood stem cells & 21 & 753 & \textsc{Out of Memory} & 5114 \\
  Embryonic (66\% of states) & 33 & 956 & \textsc{Out of Memory} & 3364  \\
  Embryonic (full) & 33 & 1448 & \textsc{Out of Memory} & 8709  \\
  \hline
\end{tabular}
  \caption{\label{fig:performance}Performance of direct encoding and compositional algorithm on example data
    sets.}
\end{center}
   \vspace*{-8mm}
\end{figure}

An implementation of our algorithm, which is written in F\# and uses
Z3 as the satisfiability solver, 
is available at \url{https://github.com/swoodhouse/SCNS-Toolkit}.
In Figure~\ref{fig:performance} we present experimental results from running our implementation of the direct encoding from Section~\ref{sec:direct} and compositional algorithm on four data sets: the small synthetic data set from Section~\ref{sec:example}, the large embryonic experimental data set from Section~\ref{sec:motivation}, and a second experimental data set covering blood stem cells. We also show results from rerunning on the embryonic data set with a third of states removed. All experiments were performed on an Intel Core i5 @ 1.70GHz with 8GB of RAM, using
a single thread.

While the direct encoding synthesised a matching Boolean network on the small synthetic data set faster than our compositional algorithm, it cannot scale to the real experimental data sets, quickly running out of memory. The compositional algorithm, on the other hand, can scale to handle real data sets of the sort produced by our experimental collaborators. All experiments terminated within a few hours, when running on a single thread.
The compositional algorithm can easily be parallelised over variables, which would further increase its efficiency.

\section{Application to the Experimental Dataset}
We now return to the experimental data set introduced in
Section~\ref{sec:motivation}.

Recall that cell measurements were taken from four sequential
developmental time points, and that the state graph resulting from
discretisation of the data (Figure~\ref{fig:chicken}) exhibited a
clear separation between the earliest developmental time point (states
coloured blue) and the latest (states coloured red). We applied our synthesis
technique to this data, taking the initial states to be the states from the first
time point, and the final states to be the states from the latest time point.
For complete details, we direct the reader to~\cite{Moignard2015}.

The result of the synthesis was a set of possible Boolean update
functions for each of the 33 genes, with several genes having a
uniquely identified update function. By applying standard techniques
for the analysis of Boolean networks, we found the stable state
attractors and performed computational perturbations. The synthesised
network, along with the subsequent computational analysis led to a set
of predictions which were then tested experimentally. We found that
our results were robust when performing bootstrapping, removing a
third of the data at random and rerunning the synthesis algorithm.

Our experimental collaborators were able to validate key predictions
made by our analysis. The update function for one of the genes at the
core of this network, \textit{Erg}, which directly activates many other genes, was
tested experimentally by a range of assays. Evidence was found that
the activators specified in the gene's synthesised update function (\textit{Hoxb4} and \textit{Sox17}) do
indeed activate expression of the gene, and furthermore in a fashion
consistent with the Boolean ``OR'' logic of the synthesised update
function. This could be regarded as a ``local'' validation of our model, testing two of the directed
edges in the network.

Computational perturbations to another gene at the core of the network, \textit{Sox7}, indicated that
when \textit{Sox7} was forced to always be expressed, stable states corresponding
to cells from the final developmental time point (blood progenitors) no longer exist. Cell
differentiation assays confirmed this prediction experimentally, finding that when
this gene was forced to be expressed, the number of cells which normally emerge
at this final time point is significantly reduced. This can be thought of as a ``global''
validation of our model, as it is a prediction about the 
behaviour of the whole network under a certain perturbation.

\section{Related Work}
\label{sec:related}
Previous analyses of single–-cell gene expression data have mostly been based on
statistical properties of the data viewed as a whole, such as the correlation
in the level of expression of pairs of genes~\cite{Moignard2013, Guo2013}.
Such analysis cannot recover mechanistic Boolean logic, does not infer the
direction of interactions and cannot easily distinguish direct from indirect influence.

Boolean networks were introduced by Kauffman in order to study random
models of genetic regulatory networks \cite{Kauffman1969}. They have
since been applied in a range of contexts, from modelling blood stem
and progenitor differentiation \cite{Bonzanni2013, Krumsiek2011}, to
the yeast apoptosis network \cite{Kazemzadeh2012}, to the
network regulating pluripotency in embryonic stem cells
\cite{Peterson2013}. BDD-based algorithms for state-space exploration
and finding attractors of Boolean networks have been 
introduced \cite{Garg2008, Zheng2013}.

Synthesis is the problem of producing programs or designs from their
specifications.  In recent years much progress has been made on the
usage of SAT and SMT solvers for synthesis.  Essentially, the
existence of a program that solves a certain problem is posed as a
satisfiability query.  Then, a solver tries to search for a solution
to the query, which corresponds to a program.  For example, Srivastava
\emph{et.~al.}~\cite{SrivastavaGF10,SrivastavaGF13} show that the
capabilities of SMT solvers to solve quantified queries enable the
search for conditions and code fragments that match a given
specification.  Similarly, Solar-Lezama
\emph{et.~al.}~\cite{Solar-LezamaRBE05} build a framework for writing
programs with ``holes'' and letting a search algorithm find proper
implementations for them.  The approach of reactive
synthesis~\cite{PnueliR89} is similar to ours in the type of artefact
that it produces.  However, the techniques that we are using are more
related to those explained above.  Recently, Beyene
\emph{et.~al.}~\cite{BeyeneCPR14} have shown how constraint solving 
can be used also in the context of reactive synthesis.

Synthesis has recently been applied in the context of
biology. K\"oksal \emph{et.~al.}~show how to synthesise
state-machine-like models from gene mutation experiments using a
novel 
counterexample-guided inductive synthesis (CEGIS) algorithm
\cite{Koksal2013}.  Their approach uses constraint solvers to search
for program completions that match given specifications, as explained
above.  Both the data and the type of model 
are different to those dealt with here, which called for a new approach.

Recently, there have been several applications of synthesis to Boolean networks.
Dunn \emph{et.~al.}\cite{Dunn06062014} and Xu \emph{et.~al.}\cite{xu2014construction} show how to fit an existing static, topological
regulatory network for embryonic stem cells to
gene expression data in order to obtain an executable Boolean network, under the assumption
that experimentally measured data represent stable states of the system. This
assumption may be appropriate for cell lines maintained in culture,
but it does not adapt well to developmental 
processes such as ours, where cells are transiting through intermediate states
in order to develop into a particular lineage.

Recent work of Karp and Sharan\cite{sharan2013reconstructing} shows how to synthesise Boolean networks
given a topological network and a set of perturbation experiments, by reduction to integer linear programming.
In \cite{AnalyzingGenomicLogicFunctions}, Paoletti \emph{et.~al.}~synthesise a related class of models (which incorporate timing and spatial information) from perturbation data, via reducion to SMT.
To the best of our knowledge, our approach is the first to synthesise gene regulatory network models
directly from raw gene expression data, without the need of either genetic perturbation data
or \emph{a-priori}~information about the topology of the network.

\section{Conclusions and Future Work}
\label{sec:conclusions}

We presented a technique for synthesising Boolean networks from
single--cell resolution gene-expression data.  This new and exciting
type of data allows us to consider the state of each cell separately, giving rise to
``state snapshots'', which we treat as the states of an asynchronous
Boolean network.  Our key insight is that the update functions of each
variable can be sought after separately, giving rise to reasonably
sized satisfiability queries.  We then combine the single gene update
functions by considering the flow of time included in the 
data.

We are able to reconstruct rules from a manually curated Boolean network and
produce a set of possible Boolean networks for the given experimental
data, for which no similar curated Boolean network is available.  The
discussion with biologists about this Boolean network led to a set of
predictions, which were then experimentally validated in the lab.

We are awaiting similar data from additional experiments to apply the
same technique to.  At the same time, we are considering the usage of
advanced search techniques, as used in this paper, to the analysis of
other types of high-throughput data.
Future work in the experimental domain includes the validation of more
of the links in our synthesised network, and the design of further
gene perturbation experiments motivated by the results of
computational perturbations. An interesting question for future research is
whether techniques like ours, which achieve scalability by treating
different aspects of a graph data structure seperately,
are applicable to other domains where graph--like data is generated.

\renewcommand{\abstractname}{Acknowledgements.}

\begin{abstract}
We thank B. Gottgens, V. Moignard, and A. Wilkinson for sharing with
us the biological data, discussing with us its biological
significance, and for discussions on the resulting Boolean network,
and its meaningfulness.  We thank R. Bodik, S. Srivastava and B. Hall for
helpful discussions.
\end{abstract}



{
  \small
  \bibliographystyle{abbrv}
  \bibliography{biology}
}

\end{document}